\documentclass[]{aa}  

\usepackage{graphicx}
\usepackage[svgnames]{xcolor}
\usepackage{txfonts}
\usepackage{url}

\begin{document} 

\title{Disappearance of the extended main sequence turn-off in intermediate age clusters as a consequence of magnetic braking}
\titlerunning{Disappearance of the extended main sequence turn-off}

   \author{{C. Georgy}\inst{1}
          \and
          {C. Charbonnel}\inst{1,2}
          \and
          {L. Amard}\inst{3}
          \and
          {N. Bastian}\inst{4}
           \and
          {S. Ekstr\"om}\inst{1}
          \and
          {C. Lardo}\inst{5}
          \and
          {A. Palacios}\inst{6}
          \and
          {P. Eggenberger}\inst{1}
          \and
          {I.~Cabrera-Ziri}\inst{7}\thanks{Hubble Fellow}
          \and 
          {F. Gallet}\inst{8} 
          \and 
          {N. Lagarde}\inst{9}
          }

   \institute{Department of Astronomy, University of Geneva, Chemin des Maillettes 51, 1290 Versoix, Switzerland\\
              \email{cyril.georgy@unige.ch}
         \and
            IRAP, UMR 5277 CNRS and Universit\'e de Toulouse, 14 Av. E. Belin, 31400 Toulouse, France
            \and
                        University of Exeter, Department of Physics \& Astronomy, Stoker Road, Devon, Exeter, EX4 4QL, UK
         \and
            Astrophysics Research Institute, Liverpool John Moores University, 146 Brownlow Hill, Liverpool L3 5RF, UK
         \and
            Laboratoire d'Astrophysique, Ecole Polytechnique F\'ed\'erale de Lausanne (EPFL), Observatoire, 1290 Versoix, Switzerland
         \and
            Universit\'e de Montpellier, CNRS, LUPM, CC 72, 34095 Montpellier Cedex 05, France
         \and
            Harvard-Smithsonian Center for Astrophysics, 60 Garden St., MS-19, Cambridge, MA 02138, USA
            \and
            Univ. Grenoble Alpes, CNRS, IPAG, 38000 Grenoble, France
            \and
            Institut UTINAM, CNRS UMR 6213, Univ. Bourgogne Franche-Comt\'e, OSU THETA Franche-Comt\'e-Bourgogne, Observatoire de Besan\c con, BP 1615, 25010, Besan\c con Cedex, France 
             }
    \authorrunning{Georgy et al.}

   \date{}

% \abstract{}{}{}{}{} 
% 5 {} token are mandatory
 
  \abstract
   {Extended main sequence turn-offs are features commonly found in the colour-magnitude diagrams of young and intermediate age (less than about $2\,\text{Gyr}$) massive star clusters, where the main sequence turn-off is broader than can be explained by photometric uncertainties, crowding, or binarity. Rotation is suspected to be the cause of this feature, by accumulating fast rotating stars, strongly affected by gravity darkening and rotation-induced mixing, near the main sequence turn-off. This scenario successfully reproduces the tight relation between the age and the actual extent in luminosity of the extended main sequence turn-off of observed clusters.}
   {Below a given mass (dependent on the metallicity), stars are efficiently braked early on the main sequence due to the interaction of stellar winds and the surface magnetic field, making their tracks converge towards those of non-rotating tracks in the Hertzsprung-Russell diagram. When these stars are located at the turn-off of a cluster, their slow rotation causes the extended main sequence turn-off feature to disappear. We investigate the maximal mass for which this braking occurs at different metallicities, and determine the age above which no extended main sequence turn-off is expected in clusters.}
   {We used two sets of stellar models (computed with two different stellar evolution codes: STAREVOL and the Geneva stellar evolution code) including the effects of rotation and magnetic braking, at three different metallicities. We implemented them in the \textsc{Syclist} toolbox to compute isochrones and then determined the extent of the extended main sequence turn-off at different ages.}
   {Our models predict that the extended main sequence turn-off phenomenon disappears at ages older than about $2\,\text{Gyr}$. There is a trend with the metallicity, the age at which the disappearance occurs becoming older at higher metallicity. These results are robust between the two codes used in this work, despite some differences in the input physics and in particular in the detailed description of rotation-induced internal processes and of angular momentum extraction by stellar winds.}
   {Comparing our results with clusters in the Large Magellanic Cloud and Galaxy shows a very good fit to the observations. This strengthens the rotation scenario to explain the cause of the extended main sequence turn-off phenomenon.}
   
   \keywords{Stars: evolution --  Stars: Hertzsprung-Russell and C-M diagrams -- Stars: rotation -- Galaxies: star clusters: general
               }

   \maketitle
%
%-------------------------------------------------------------------

%%%%%%%%%%%%%%%%%%%%%%%%%%%%%%%%%%%%%%%%%%%%%%%%%%%%%%%%%%%%%%%
\section{Introduction}
%%%%%%%%%%%%%%%%%%%%%%%%%%%%%%%%%%%%%%%%%%%%%%%%%%%%%%%%%%%%%%%

During the past decade, exquisite observations with the Hubble Space Telescope (HST) revealed unexpected features in the colour-magnitude diagram (CMD) of young massive star clusters \citep[YMC; see e.g.][]{Mackey2007a}. For instance, these clusters exhibit an unusual main sequence turn-off (MSTO), which is more extended than expected from a single stellar population (SSP). One possible explanation for this feature is that YMC have undergone several episodes of star formation, and are thus composed of stellar populations of different ages \citep[e.g.][]{Correnti2014a,Goudfrooij2015a}.

On the other hand, rotation has been suggested as a possible cause of the extended MSTO (eMSTO) phenomenon \citep{Bastian2009a,Yang2013a,Brandt2015a,Milone2018a}. One strong prediction of this scenario is that if the extension of the MSTO due to rotation is interpreted as an age spread, then the inferred age spread should increase as a function of the age of the cluster \citep{Niederhofer2015a}. This is fully consistent with the observational data \citep[][]{Bastian2016a,Bastian2018a}. In fact, it is now clear that the position of a star at the MSTO is dependent on the stellar $V\sin(i)$, in agreement with the rotational scenario \citep{Dupree2017a,Bastian2018a,Marino2018a,Kamann2018a}. In case this scenario is correct, we expect the eMSTO phenomenon to disappear for clusters old enough \citep{Brandt2015a}, so that stars lying at the MSTO have a mass small enough to have undergone significant magnetic braking early on the MS. This braking is produced by the interaction of the stellar winds with the surface magnetic field generated by a dynamo process in the external convection zone of the star \citep[e.g.][]{Sch62,WD67,Kawaler88,Mattetal2015}. Hence, clusters above a certain age limit should display a narrow (effectively a SSP) MSTO, consistent with recent observations \citep{Martocchia2018a}. \citet{Brandt2015a} suggested this should occur above $2\,\text{Gyr}$ at LMC metallicity, but could not provide better constraints, because the stellar model grids they used \citep{Georgy2013a} do not contain models for sufficiently low  masses.

Here we explore this age limit within the rotational scenario, using a grid of low-mass star models computed for a large range of metallicities and initial rotation rates. In Sect.~\ref{RotProperties}, we recall the effects of rotation on stellar evolution, which are relevant in the framework of this scenario. The properties of the new models are discussed in Sect.~\ref{Models}. In Sect.~\ref{eMSTO}, we use these models and a set of models from previous grids to explore the age beyond which the eMSTO feature is expected to disappear, and compare our findings to observations from the LMC. We conclude in Sect.~\ref{Conclusions}.

%%%%%%%%%%%%%%%%%%%%%%%%%%%%%%%%%%%%%%%%%%%%%%%%%%%%%%%%%%%%%%%
\section{Main effects of rotation on stellar models and isochrones} \label{RotProperties}
%%%%%%%%%%%%%%%%%%%%%%%%%%%%%%%%%%%%%%%%%%%%%%%%%%%%%%%%%%%%%%%

The effect of rotation on stellar models has been extensively discussed in the literature \citep[see e.g. the reviews by][]{Maeder2000b,Maeder2012a,Palacios2013}. We recall here only the most relevant points in the context of this study for main sequence (MS) stars:
\begin{itemize}
\item \textbf{\textit{Hydrostatic effects:}} By adding a centrifugal support force, a rotating star behaves as a non-rotating star of lower mass. This means that  on the zero-age main sequence (ZAMS), it will have a lower effective temperature in the Hertzsprung-Russell Diagram (HRD) due to the additional centrifugal support force \citep[e.g.][]{Brott2011a,Ekstrom2012a}.
\item \textbf{\textit{Mixing effects:}} Depending on the adopted prescriptions for rotation-induced transport processes (i.e., turbulence and meridional circulation), the mixing of chemical species at the convective core boundary is more or less efficient. In case of inefficient mixing, the tracks of rotating models in the HRD are almost parallel to the non-rotating tracks, with only the hydrostatic effects discussed above. In case of efficient mixing, the core is refuelled with hydrogen and its mass increases \citep[e.g.][]{Heger_Langer_00}. The track of the rotating model crosses the non-rotating one and exhibits a higher luminosity at the end of the MS \citep{Meynet2013a}. Moreover, the MS duration is increased with respect to non-rotating models \citep{Ekstrom2012a,Lagarde_etal12}.
\item \textbf{\textit{Viewing angle:}} Fast rotating stars have a non-uniform brightness and temperature \citep[e.g.][]{vonZeipel1924a,EspinosaLara2011a}. Rotating stars seen pole-on appear hotter and brighter, while those seen equator-on are cooler and fainter. This affects the position of the star in the HRD \citep{Georgy2014b}.
\end{itemize}
All these effects play an important role in the morphology of the HRDs or CMDs of stellar clusters, and have to be accounted for when discussing the aspects of the MSTO.

%%%%%%%%%%%%%%%%%%%%%%%%%%%%%%%%%%%%%%%%%%%%%%%%%%%%%%%%%%%%%%%
\section{Stellar model predictions} \label{Models}
%%%%%%%%%%%%%%%%%%%%%%%%%%%%%%%%%%%%%%%%%%%%%%%%%%%%%%%%%%%%%%%

\subsection{Prescriptions for rotation and magnetic braking}\label{RotBraking}

We use a new grid of standard and rotating stellar models computed by Amard et al. (2018, in prep.) with the code STAREVOL for different metallicities (here we use $\left[\text{Fe}/\text{H}\right] = -1.0$, $-0.3$, and $0.0$, roughly corresponding to the SMC, LMC, and Galactic metallicity, respectively), and for masses between $1$ and $2\,M_\sun$. We refer to the original paper for information on the input physics (initial chemical mixtures, nuclear network and reaction rates, equation of state, opacities, treatment of convection, mass loss, and model atmosphere), and to \citet{DecressinMathis2009} and \citet{2018arXiv180801814M} for the description of the transport of angular momentum and chemicals by meridional circulation and shear turbulence in the interior of the models including rotation. The extraction of angular momentum at the stellar surface by magnetised winds follows the prescription of \citet{Mattetal2015} calibrated for a $1\,M_\sun$, $Z_\sun$ model to reproduce the mean solar rotation rate at the age of the Sun. The corresponding torque applied to the models is a function of the stellar mass, radius, surface  angular velocity ($\Omega$), and convective turnover timescale ($\tau_\text{c}$). Its efficiency mostly depends on the convective Rossby number ($\text{Ro} = \frac{P_\text{rot}}{\tau_\text{c}}$, with $P_\text{rot} = 2\pi/\Omega$), which is a good proxy for the dynamo efficiency \citep[e.g.][ and references therein]{Charbonneletal17}. No core overshooting is considered for the STAREVOL models used here.

Each model ($M$; $\left[\text{Fe}/\text{H}\right]$; rotation rate) of the grid is computed from the pre-main sequence (PMS) to the base of the red giant branch. Three different initial rotation velocities are adopted for the grid. They correspond to a combination of initial rotation period and disc-coupling duration at the beginning of the PMS, which is chosen to reproduce the observed dispersion of the rotation periods of low-mass stars in young Galactic star clusters and associations with ages between $1\,\text{Myr}$ and $2.5\,\text{Gyr}$ (see \citealt{Amard2016} and 2018 for details, and \citealt{Bouvier2014a} for references of the photometric surveys). The same combination is kept over the entire mass and metallicity range. Over the entire mass range considered, the surface angular velocity increases along the PMS after the end of the disc-coupling phase as a result of stellar contraction. In the case of low-mass stars, it reaches a maximum value when the model approaches the ZAMS, before eventually decreasing on the MS at a rate that depends on the actual magnetic braking (see below). For this study, we select the ``median rotation'' models, which are associated to the 50th percentiles of the statistical sample in each cluster.

\subsection{Model predictions at LMC metallicity}\label{section_predLMC}

\begin{figure*}
\centering
\includegraphics[width=.95\textwidth]{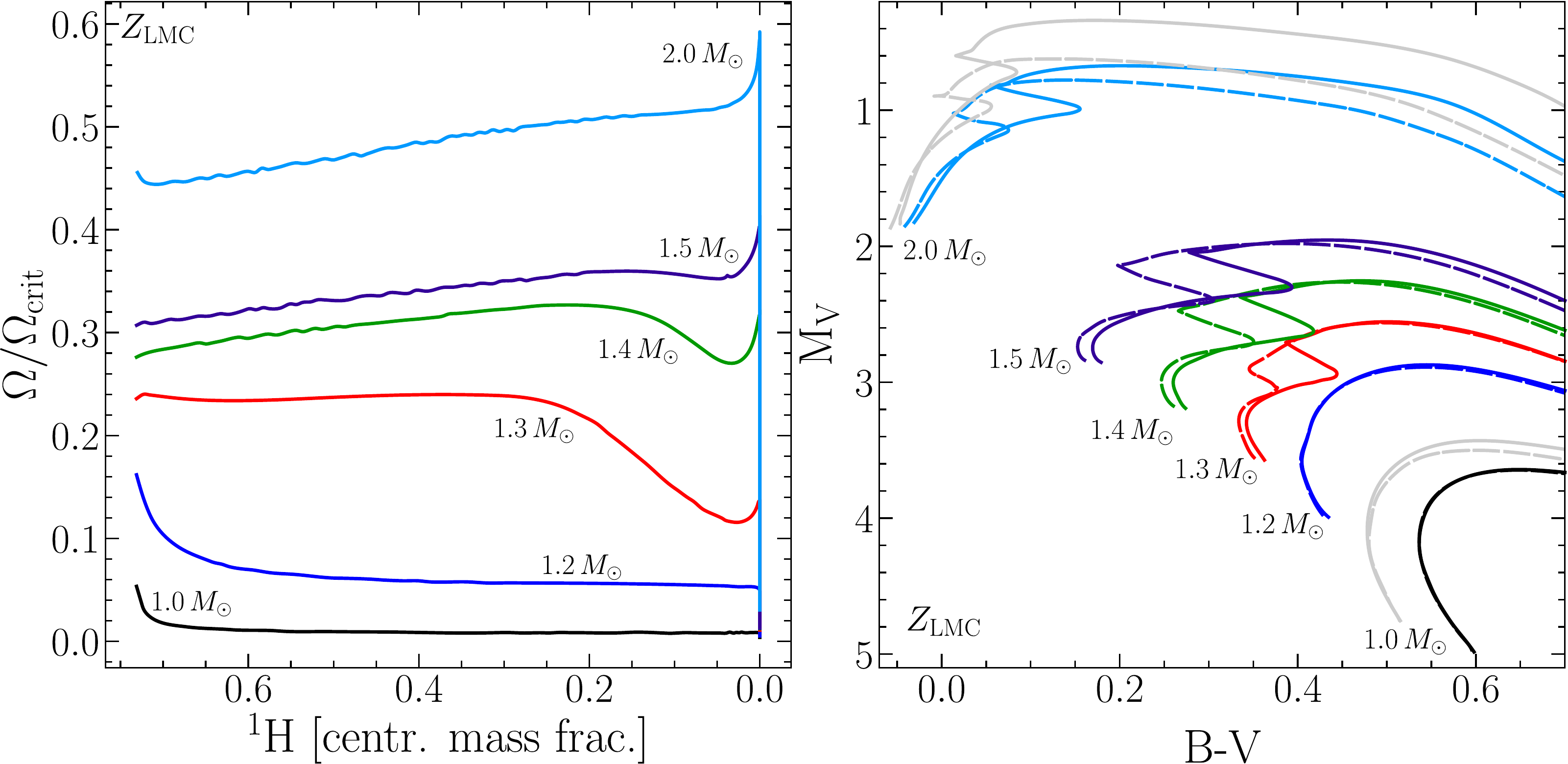}
\caption{Left panel: Evolution of the ratio of the surface angular velocity to the critical velocity for STAREVOL models with $\left[\text{Fe}/\text{H}\right] = -0.3$ relevant for LMC metallicity (this corresponds to a metal mass fraction $Z = 0.008$ for the adopted initial chemical mixture). Central hydrogen mass fraction (in abscissa) is a proxy for time evolution along the MS (left to right). For each mass (colour-labelled), model predictions are shown for the median rotators of the grid of Amard et al. (2018, in prep.). The transition is clearly visible between models that are braked early on the MS by the coupling of the surface magnetic field and the stellar winds ($\leq 1.2\,M_\sun$), and models that are never braked ($\geq 1.5\,M_\sun$). Right panel: Evolution tracks in the CMD for stellar models of different masses (see label near the tracks, the colour code for the STAREVOL models is the same as in left panel), at the metallicity of the LMC. Standard models (no rotation, dashed lines) are compared with the rotating models (solid lines). For comparison purposes, we also indicate the tracks of $1\,M_\sun$ and $2\,M_\sun$ models computed with the Geneva code (GENEC), shown in grey \citep[Eggenberger et al. in prep., same physics as in][]{Ekstrom2012a} at a metallicity $Z=0.006$, corresponding to $\left[\text{Fe}/\text{H}\right] = -0.376$. The metallicity and initial mixture of the two sets of models are not exactly the same, explaining the slight shift of the tracks from the ZAMS on.}
\label{OmOmcrit_LMC}
\end{figure*}

To illustrate the general trends, we first focus on model predictions at LMC metallicity ($\left[\text{Fe}/\text{H}\right]=-0.3$). The MS evolution of the ratio between the surface angular velocity and its critical value ($\Omega_\text{crit} = \sqrt{\frac{GM}{R_\text{e, crit}^3}}$, with $R_\text{e, crit}$ the equatorial radius at the critical velocity; \citealt{Maeder2009a}) is shown in Fig.~\ref{OmOmcrit_LMC} (left panel) for a few initial masses, as a function of the central hydrogen fraction that is used as a proxy for time between the ZAMS and the MSTO. Depending on the stellar mass, surface rotation during the MS follows three different regimes. For the lowest masses ($1.2\,M_\sun$ and below) that have an extended convective envelope (CE) with relatively long convective turnover timescales, hence low Rossby numbers, surface rotation is driven by the fast extraction of angular momentum by the strong magnetic torque. In this case, magnetic braking is very efficient early on the MS, and one does not expect to find fast rotating stars near the MSTO. At the other extreme, more massive stars ($1.5\,M_\sun$ and above) that have extremely thin (or no) CE do not undergo significant magnetic braking. The ratio between the surface angular velocity and its critical value increases from the ZAMS to the end of the MS. In this regime, the existence of fast rotating stars near the MSTO is possible. Finally, in the intermediate regime ($1.3-1.4\,M_\sun$), some braking can occur later on the MS, when the star becomes cool enough to develop a significant CE.

In terms of tracks in the HRD or CMD (Fig.~\ref{OmOmcrit_LMC}, right panel), the relevant parameter is the efficiency of the internal mixing. With the prescription for anisotropic turbulence by \citet{2018arXiv180801814M} used in Amard et al. (2018, in prep.), the models that are efficiently braked early on the MS ($M\lesssim 1.2\,M_\sun$) develop a very weak mixing in the central regions of the star, making the tracks almost indistinguishable from the standard ones. More massive models, which are not braked (or later on the MS), present however the usual features of rotating models, with a MSTO being shifted towards lower effective temperature and higher luminosity. Internal mixing contributes also to an increase of the duration of the MS, by refuelling the core with fresh hydrogen. This is illustrated in Fig.~\ref{LifeTime_LMC}, where we show the ratio of the MS duration of rotating versus non-rotating models. For the STAREVOL models (solid blue line) we use here, this ratio is very close to $1$ for models that are braked efficiently, and then increases to a plateau value of about $1.18$ for more massive models. 

In summary, there is a clear transition between the cool, low-mass stars with extended CE that undergo strong magnetic braking on the early MS and for which rotation does not lead neither to a change in the HRD/CMD tracks, nor to an increase of the duration of the MS, and the hotter, more massive models with extremely thin or no CE that are not braked by stellar winds and for which rotation induces a notable modification of the tracks and a longer MS lifetime.

\subsection{Metallicity dependence}

Magnetic braking on the early MS strongly depends on the characteristics of the stellar CE (through $\tau_\text{c}$) as described above, and it operates on very short timescales  for stars with $T_\text{eff}$ on the ZAMS cooler than $\sim 7000-6500\,\text{K}$. The effective temperature of a MS star is mainly a function of its mass and metallicity (and eventually rotation). At a given $T_\text{eff}$, we find a slightly more massive and less luminous star if the metallicity is higher \citep[opacity effect, see also][]{Schaller1992a}. However, the convective and magnetic characteristics are very similar \citep[e.g.][]{TalonCharbonnel2004,Charbonneletal17}. As a consequence, the  mass limit at the transition between the braking and no braking regimes is shifted towards higher mass when metallicity increases. This is illustrated in Fig.~\ref{LifeTime_LMC}, where the increase of the MS lifetime due to rotation in STAREVOL models is progressively increased towards higher mass when metallicity is decreasing.

\subsection{Comparison with Geneva models}

\begin{table*}
\caption{Summary of the main similarities and differences between STAREVOL and GENEC models, which are relevant for this paper.}
\label{TabDiff}
\begin{center}
\begin{tabular}{c|cc}
 & STAREVOL & GENEC\\
\hline
rotation & \citet{Zahn1992,Maeder1998a} & \citet{Zahn1992,Maeder1998a}\\
turbulence ($D_\text{h}$, $D_\text{v}$) & \citet{2018arXiv180801814M,Zahn1992} & \citet{Zahn1992,Maeder1997a}\\
atomic diffusion & no & yes\\
magnetic braking & \citet{Mattetal2015} & \citet{Krishnamurthi1997a}\\
convection & Schwarzschild & Schwarzschild\\
 & & $0.1\,H_P$ for $M\geq1.7\,M_\sun$\\
overshooting & no & $0.05\,H_P$ for $1.25 \leq M <1.7\,M_\sun$\\
 & & $0$ for $ M <1.25\,M_\sun$\\
\end{tabular}
\end{center}
\end{table*}

To check the robustness of our results, we also use in this work grids of models computed with the Geneva stellar evolution code (GENEC) at similar metallicities \citep[even if slightly different, see][Eggenberger et al. in prep.]{Ekstrom2012a,Georgy2013b}. The main differences between STAREVOL and GENEC models are (also summarised in Table~\ref{TabDiff}):
\begin{itemize}
\item Rotation is treated in the same general framework \citep{Zahn1992,Maeder1998a}. However, the detailed implementation of the horizontal and shear diffusion coefficient as well as the assumed prescriptions for turbulence are different.
\item GENEC models include a small amount of overshooting ($0.05H_\text{P}$, with $H_\text{P}$ the pressure scale height, between $1.25$ and $1.5\,M_\sun$; $0.1H_\text{P}$ above $1.5\,M_\sun$), when STAREVOL models used here have no overshooting included.
\item In GENEC models, braking of the stellar surface by magnetised winds is applied for stars with an external convective envelope ($M < 1.7\,M_\sun$ at solar metallicity). The braking law of \citet{Krishnamurthi1997a} is adopted with the braking constant being calibrated so that the $1\,M_\sun$ rotating model reproduces the solar surface rotational velocity after $4.57\,\text{Gyr}$.
\item In GENEC grids, the PMS is not fully computed. The initial velocity of the rotating models on the ZAMS is the following (for all metallicities): for models with initial mass $M\geq1.7\,M_\sun$, it corresponds to $40\%$ of the critical velocity (about $200\,\text{km}\,\text{s}^{-1}$ for the $2\,M_\sun$ model at the LMC metallicty). The $1.5\,M_\sun$ starts at $150\,\text{km}\,\text{s}^{-1}$. The $1.25$ and $1.35\,M_\sun$ starts at $100\,\text{km}\,\text{s}^{-1}$. Lower mass models start at $50\,\text{km}\,\text{s}^{-1}$. For comparison, the corresponding STAREVOL models with $[\text{Fe}/\text{H}] = -0.3$ have a rotation velocity at the arrival on the MS of about $19\,\text{km}\,\text{s}^{-1}$ for the $1\,M_\sun$ model, $110\,\text{km}\,\text{s}^{-1}$ for the $1.5\,M_\sun$ model, and $180\,\text{km}\,\text{s}^{-1}$ for the $2\,M_\sun$ one.
\item In GENEC models, the effects of atomic diffusion due to concentration and thermal gradients are taken into account \citep[See Sect.~3 of][for more details]{Eggenberger2008a}.
\end{itemize}

\begin{figure}
\centering
\includegraphics[width=.45\textwidth]{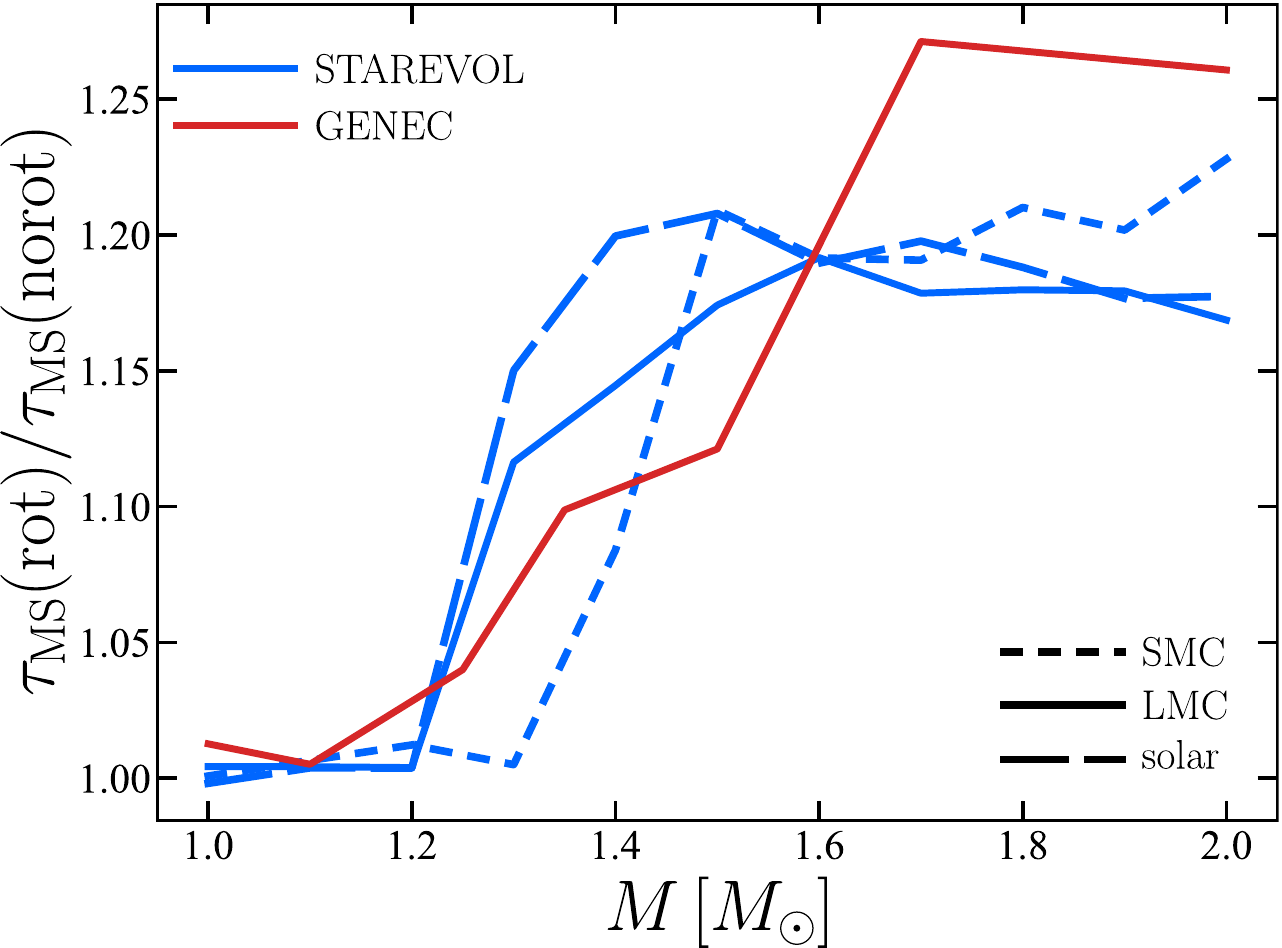}
\caption{Ratio of the MS duration of rotating to non-rotating models from both STAREVOL (blue) and GENEC (red) models in the mass range $1-2\,M_\sun$. GENEC models have the metallicity of the LMC. STAREVOL models have SMC (long dashed), LMC (solid), and solar (short dashed) metallicity.}
\label{LifeTime_LMC}
\end{figure}

As mentioned previously, STAREVOL models of Amard et al. (2018, in prep.) that include the \citet{2018arXiv180801814M} prescription for anisotropic turbulence have a quite inefficient mixing when the star is braked early on the MS. In addition to the disappearance of the hydrostatic effects, it makes the tracks of the lowest masses considered converge towards the non-rotating ones, and the ratio of the rotating to non-rotating MS duration close to 1. On the contrary, the mixing inside GENEC models is more efficient, with the following consequences. For low-mass stars, the tracks of the rotating models deviate from the non-rotating ones (see the track of the $1\,M_\sun$ model shown in light grey in Fig.~\ref{OmOmcrit_LMC}, right panel). It is related to the inclusion of atomic diffusion in GENEC models, which impacts the non-rotating tracks. Indeed, the main impact of rotational mixing in low-mass stars is to counteract the effects of atomic diffusion \citep[see][for more details]{Eggenberger2010a}. The effect on the lifetime is, however, small, the ratio of the MS duration of rotating and non-rotating models remaining close to 1 for models with $M\leq1.25\,M_\sun$ (see the red curve in Fig.~\ref{LifeTime_LMC}). The stronger mixing also impacts more massive models, making the tracks of the rotating models more vertical in the CMD, and reaching higher luminosities ($2\,M_\sun$ model in grey, Fig.~\ref{OmOmcrit_LMC}, right panel). The MS duration ratio of rotating to non-rotating models also reaches a plateau, however slightly higher than the STAREVOL models (about $25\%$).

By affecting both lifetimes and HRD/CMD tracks, these differences will impact the shape of the isochrones. It is thus particularly interesting to check how both sets of models behave near the MSTO to validate the robustness of our results with respect to changes in the physical ingredients included in the computations.

\section{Isochrones and implications on the extended main sequence turn-off}\label{eMSTO}

\subsection{Impact of rotation on the isochrones}

\begin{figure*}
\centering
\includegraphics[width=.95\textwidth]{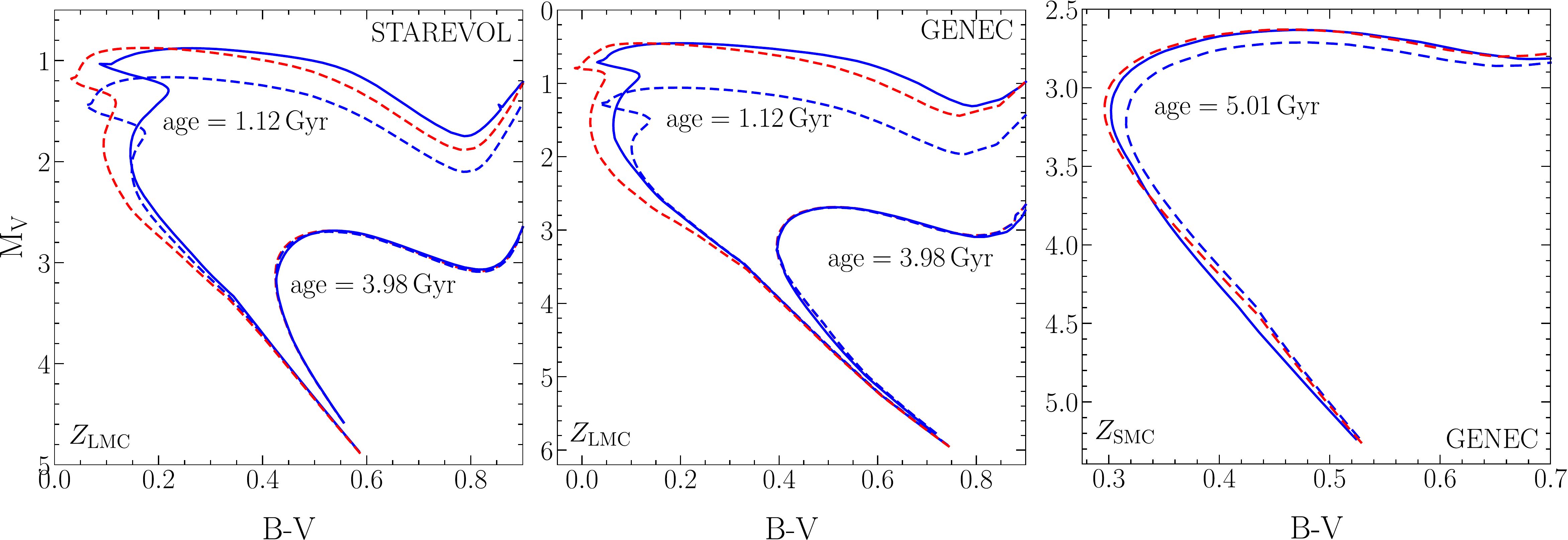}
\caption{Isochrones at about the LMC metallicity computed with STAREVOL ($Z=0.008$, left panel) and GENEC ($Z=0.006$, central panel), and at the SMC metallicity with GENEC ($Z=0.002$, right panel). The blue curves are the isochrones for the ages indicated on the plot for rotating (solid), and non-rotating (dashed) models. Left and central panel: one set of isochrones illustrates an age where a large eMSTO is expected ($1.12\,\text{Gyr}$), and the other set a much older age, where the isochrones are almost indistinguishable ($3.98\,\text{Gyr}$). The red dashed curves represent the non-rotating isochrones that best match the turn-off luminosity of the rotating model (see text). These isochrones correspond to ages of $0.92$ and $3.94\,\text{Gyr}$ for STAREVOL models and $0.74$ and $3.97\,\text{Gyr}$ for the GENEC models at the LMC metallicity. The small differences in term of colour and magnitude between both sets of models is due to the slightly different initial metallicity and microphysics included in the modelling. Right panel: Same as the other panels, at an age of $5.01\,\text{Gyr}$ for GENEC rotating models. Here, the age of the dashed red (non-rotating) isochrone is $4.66\,\text{Gyr}$}
\label{Isochrones}
\end{figure*}

In our previous works \citep{Niederhofer2015a,Bastian2016a} we have investigated the effect of rotation on the MSTO of clusters younger than $1\,\text{Gyr}$ (MSTO masses higher than about $1.7\,M_\sun$). We have shown that the eMSTO general aspect can be explained by a population of stars with different initial rotation velocities.\footnote{\footnotesize{The exact distribution of stars inside the eMSTO is more difficult to reproduce \citep{Goudfrooij2017a}. However, a detailed study of the morphology of the eMSTO by using a complete \textsc{Syclist} (\url{https://unige.ch/sciences/astro/evolution/fr/base-de-donnees/syclist/}) synthetic cluster and comparing with observed CMDs of clusters would require a much more sophisticated method to compute the colours of near-critically rotating stars than the one currently implemented in this toolbox \citep{Georgy2014b}.}} In this framework, the eMSTO is produced by a combination of the effects that we have just illustrated and that we briefly recall. First, rotation enlarges the MS in the HRD by shifting the tracks towards higher luminosity and lower $T_\text{eff}$. However, the longer MS lifetime of the rotating models mitigates this effect, making the temperature of the MSTO of the isochrones of rotating models similar to that of non-rotating models \citep[see Fig.~\ref{Isochrones} and][]{Girardi2011a}. This effect alone is therefore not able to explain the eMSTO feature. Second, fast rotating models have a longer lifetime than slow rotating models of a similar mass and metallicity. At a given age, there is therefore an accumulation of fast rotating stars (still on the MS) near the MSTO, while their slowly rotating counterpart have already evolved through the Hertzsprung gap. Third, due to gravity darkening, fast rotating stars appear different depending on the viewing angle. A near critically rotating star seen pole-on will appear hotter and brighter than shown in Fig.~\ref{OmOmcrit_LMC}, whereas the same star seen equator-on will appear cooler and dimmer \citep{EspinosaLara2011a,Georgy2014b}. Since eMSTO cannot be reproduced by a classical non-rotating single stellar population \citep[e.g.][]{Mackey2007a}, and assuming that all the stars of the cluster have the same age, its cause should be explained by a physical effect able to act differently in stars of the same age and mass. In this framework, rotation appears to be a natural candidate. Other possibilities such as a variable overshooting or other effects have been proposed in the literature \citep[e.g.][]{Yang2017a}, but rely so far on sparse observational evidences and lack theoretical explanations.

The combined impact of the first two effects above on the position of isochrones in the CMD for the LMC metallicity is illustrated by the blue tracks on Fig.~\ref{Isochrones} for STAREVOL (left panel) and GENEC models (right panel). The isochrones were computed by processing the two sets of models in the \textsc{Syclist} toolbox following the same procedure as in \citet{Georgy2014b}. At ages of about $1\,\text{Gyr}$ or younger, the isochrone is clearly shifted towards higher luminosity when comparing the rotating (solid line) and non-rotating (dashed line) models. This is not surprising as the turn-off mass at this age ($\sim1.8\,M_\sun$) corresponds to stars that are never braked during the MS phase (see Sect.~\ref{section_predLMC}). At older age (typically $4\,\text{Gyr}$), the isochrones for rotating and non-rotating models are very similar and almost indistinguishable. This is true for both STAREVOL and GENEC model sets. It is easily understandable for STAREVOL models, since the tracks and lifetime are extremely similar for stars that are efficiently braked during the MS. For GENEC models, where this convergence is less visible, the small shift in the CMD between the rotating and non-rotating tracks is compensated by the slightly different MS lifetime (see Figs.~\ref{OmOmcrit_LMC}, right panel, and \ref{LifeTime_LMC}).

\subsection{Extension of the MSTO}

For each of the three metallicities considered here, we computed the equivalent age spread $\Delta t_\text{eMSTO}$ that would be required to reproduce the same luminosity shift between the rotating and non-rotating isochrones at different ages, following exactly the same procedure as in \citet{Niederhofer2015a}. We used their minimum $\text{M}_\text{V}$ method in our case, since it is the only feature appearing on all the isochrones, including the older ones. The results are illustrated by the red curves in Fig.~\ref{Isochrones}, which correspond to the non-rotating isochrone that best matches the magnitude of the rotating one. The age difference between the rotating isochrone and this best-matching, non-rotating one determines the equivalent age spread $\Delta t_\text{eMSTO}$.

The results are shown in Fig.~\ref{eMSTOexpect} for both sets of models. The GENEC models include the tracks of Eggenberger et al. (in prep.), \citet{Ekstrom2012a}, and \citet{Georgy2013b}, hence they cover a much broader mass range than the STAREVOL ones, explaining the extension of the curve at lower ages. The results generally agree very well: $\Delta t_\text{eMSTO}$ increases linearly with time until an age of about $1-2\,\text{Gyr}$, depending on the metallicity. At older ages, all the curves show an abrupt decrease, due to the arrival at the MSTO of models previously efficiently braked during the MS. The age at which this drop occurs increases with increasing metallicity. This is the result of the combined effect of the metallicity dependent evolution of both  the mass threshold for efficient braking on the MS and the MS duration. The mass limit below which a star is efficiently braked increases with metallicity (since it occurs at roughly a constant effective temperature at any metallicity). This means that the typical mass at the top of the $\Delta t_\text{eMSTO}$ curve is higher at higher metallicity. We could conclude that it implies that the drop occurs at a younger age at higher metallicity (due to the shorter MS duration of more massive stars). However, the MS duration increases strongly with increasing metallicity in the mass range we consider here \citep[e.g.][see their Fig.~5]{Georgy2013b}. The latter effect is stronger than the former, making the drop in $\Delta t_\text{eMSTO}$ occur at an older age at higher metallicity.

Figure~\ref{eMSTOexpect} also displays information about the rotation of the models at the MSTO: a thick line means that the models are never braked during the MS and can thus be fast rotating at the MSTO. On the other hand, a thin line means that the models are efficiently braked, and are thus slowly rotating at the MSTO. The dashed line represents the intermediate cases. According to the rotational scenario for explaining the eMSTO feature, both a large equivalent age-spread and the possibility of having fast rotating stars (and thus, gravity darkening and viewing angle effect) at the MSTO are required. It thus means that eMSTO should be observable up to an age of about $2\,\text{Gyr}$ (slightly higher at lower metallicity), and then disappear at older ages. Some of our curves seem to re-increase for ages older than $3-4\,\text{Gyr}$ (for example, the GENEC model curve at $Z=0.002$). The reason for this is illustrated by the right panel of Fig.~\ref{Isochrones}. In some cases, there is a persistent difference in the isochrones between the rotating and non-rotating models, making the fit of a younger non-rotating isochrone (the red dashed curve shows it for $\text{age} = 4.66\,\text{Gyr}$) require a somewhat younger age to match the magnitude. However, we see clearly by looking at the isochrones that is by far not comparable to the large spread in magnitude we observe at younger ages (see left and central panel). Moreover, at such an advanced age, the models are all slowly rotating when reaching the MSTO, so that we do not expect this to appear as an eMSTO in the CMD.

As can be seen in Fig.~\ref{eMSTOexpect}, the exact age at which the eMSTO phenomenon disappears is model-dependent. It depends on the exact implementation of rotation, convection, and overshooting, and initial chemical mixture of the models, which all impact the predicted lifetime of a stellar model, and the expected stellar surface properties. It is not the aim of this paper to provide exact values for this age. However, it predicts that the eMSTO phenomenon should disappear in sufficiently old  clusters. Moreover, the metallicity trend seems to be robust based on two independent grids of models.

\section{Comparison with observations}\label{Observations}

\begin{figure}
\centering
\includegraphics[width=.5\textwidth]{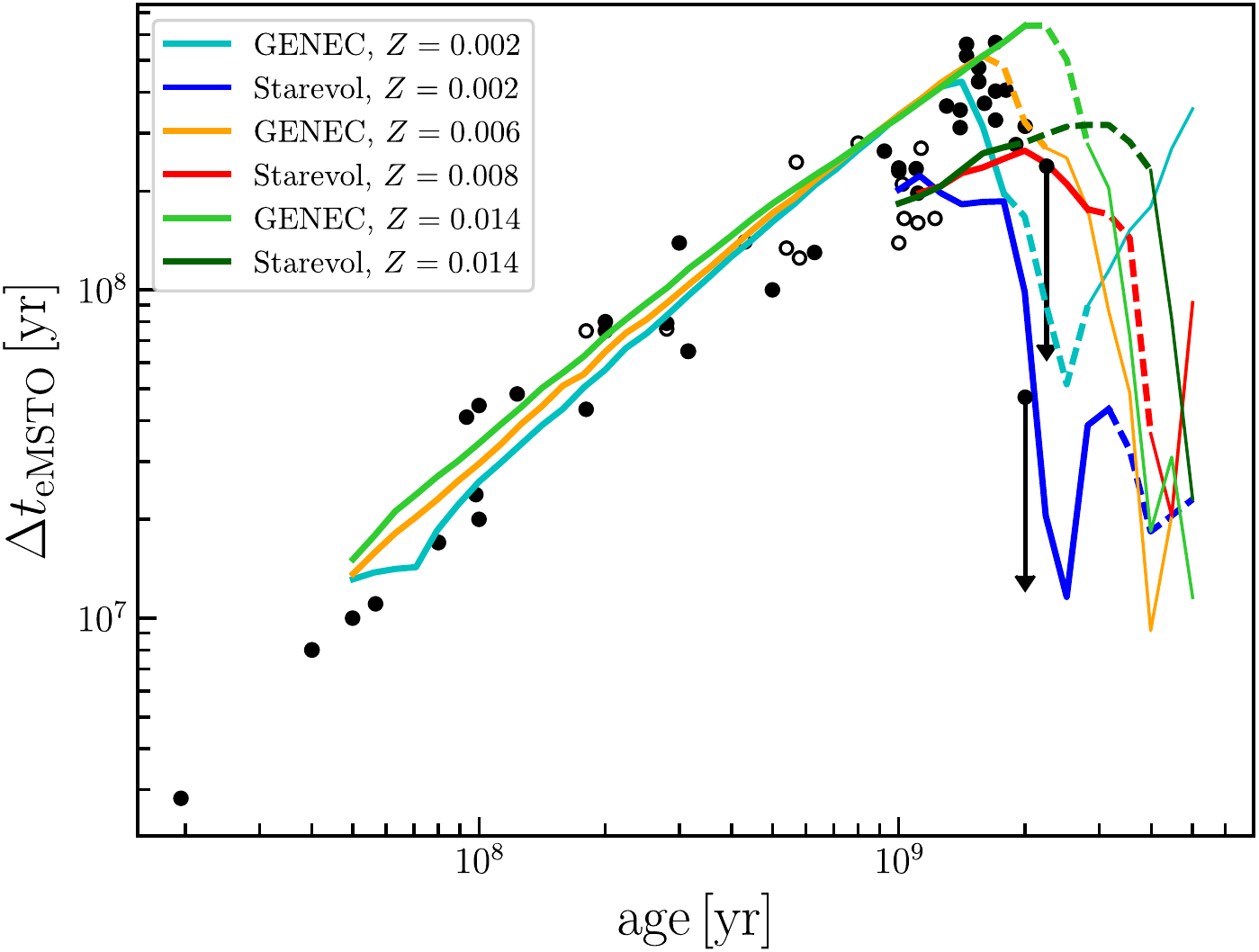}
\caption{Clusters for which an eMSTO has been observed in the LMC \citep[from][filled circles]{Goudfrooij2014a,Milone2015a,Niederhofer2015b,Bastian2016a,Milone2017a,Goudfrooij2017a,Martocchia2018a,Milone2018a} and in the Galaxy \citep[][open circles]{Bastian2018a,Cordoni2018a}. Down arrows correspond to upper limits. Coloured curves are the results in term of equivalent age spread from our simulations at three metallicities for both stellar evolution codes (see text for full description of line types).}
\label{eMSTOexpect}
\end{figure}

In Fig.~\ref{eMSTOexpect}, the inferred age spread (or equivalently, the presence of an eMSTO) of LMC clusters is plotted as a function of the cluster age \citep[data from][]{Goudfrooij2014a,Milone2015a,Niederhofer2015b,Bastian2016a,Milone2017a,Goudfrooij2017a,Martocchia2018a,Milone2018a} and Galactic clusters \citep{Bastian2018a,Cordoni2018a}. All clusters younger than about $2\,\text{Gyr}$ exhibit an eMSTO. Beyond this age, this feature seems to disappear. We draw attention to NGC 1978 (leftmost point with an upper limit), which is a massive ($\sim 3\cdot 10^5\,M_\sun$) cluster with an age of $\sim 2\,\text{Gyr}$.  \citet{Martocchia2018a} studied the MSTO width in this cluster and concluded that it was consistent with the observational uncertainties, placing an upper limit of an equivalent age spread of $\text{FWHM}=60\,\text{Myr}$. This cluster is past the peak in the distribution in Fig.~\ref{eMSTOexpect}, and suggests a rapid decline in the eMSTO spread after the peak.

The age of this transition is qualitatively consistent with the results of our modelling. However, we ask the reader to keep in mind that the cluster ages used in Fig.~\ref{eMSTOexpect} come from various sources, and are neither fully consistent within each other, nor with the stellar models we have considered in this paper. This could add a shift of the observational points along the $x$-axis. Obviously, observational data points for clusters older than $2\,\text{Gyr}$ are definitively needed to ultimately test the rotation scenario, as well as observations in different metallicity environments. In a recent paper, \citet{Goudfrooij2018a} have challenged the rotational scenario, arguing that a colour spread is seen in the CMD in clusters for stellar masses below $1.5$~M$_{\odot}$.  However, as shown in the present work, the effects of rotation are expected to continue to lower masses, potentially explaining the observations. We intend to explore, in depth, the transition where eMSTO disappears by including the new set of models presented in this work in our synthetic cluster code {\textsc{Syclist,}} taking into account the gravity darkening effects in a self-consistent way. This shall be done in a forthcoming paper.

\section{Conclusions}\label{Conclusions}

In this paper, we have explored the consequences of stellar rotation on the eMSTO feature over a broad range of ages and metallicities. At younger ages, the predictions of rotating stellar models \citep[e.g.][]{Georgy2013a,Niederhofer2015a} provide a coherent picture of the eMSTO as a function of the cluster age. For older clusters, we have shown that the eMSTO feature should disappear beyond a given age (dependent on the metallicity, $\sim2\,\text{Gyr}$ at the LMC metallicity). This behaviour is directly related to the convective envelope becoming thick enough for magnetic braking to efficiently spin down the star, thus suppressing both the effect of hydrostatic equilibrium and of chemical mixing on MS broadening. Our models show that the age at which the eMSTO feature disappears increases with increasing metallicity. These results are sustained by two sets of models computed with different codes and including slightly different prescriptions for the rotation-induced mechanisms and convection, as well as slightly different initial chemical mixtures.

Comparison with a sample of observed clusters in the LMC shows good agreement with the theoretical models. This strengthens the fact that stellar rotation is a key ingredient in causing eMSTO in young and intermediate age clusters. A more detailed study of the morphology of the eMSTO by extending the \textsc{Syclist} capabilities in modelling synthetic clusters towards lower mass is planned in a forthcoming paper.

\begin{acknowledgements}
The authors thank the referee for her or his constructive report that helped to improve this work. The authors acknowledge financial support from the Swiss National Science Foundation (SNF), the French Programme National de Physique Stellaire (PNPS) and Programme National  Cosmologie et Galaxies (PNCG) of CNRS/INSU, and the grant ANR 2011 Blanc SIMI5-6 020 01 Toupies. CC, CG, and LA thank the Equal Opportunity Office of the University of Geneva. LA also thanks the ERC (grant 682393, AWESoMe Stars). NB gratefully acknowledges financial support from the Royal Society (University Research Fellowship) as well as from the European Research Council (ERC-CoG-646928, Multi-Pop). CL acknowledges financial support from the Swiss National Science Foundation (Ambizione grant PZ00P2\_168065). F.G acknowledges financial support from the CNES fellowship and from the European Research Council (ERC) under the European Union's Horizon 2020 research and innovation programme (grant agreement No 742095; SPIDI: Star-Planets-Inner Disk-Interactions). Support for this work was provided by NASA through Hubble Fellowship grant HST-HF2-51387.001-A awarded by the Space Telescope Science Institute, which is operated by the Association of Universities for Research in Astronomy, Inc., for NASA, under contract NAS5-26555.
\end{acknowledgements}

\bibliographystyle{aa}
\bibliography{Biblio}

\begin{thebibliography}{54}
\expandafter\ifx\csname natexlab\endcsname\relax\def\natexlab#1{#1}\fi

\bibitem[{{Amard} {et~al.}(2016){Amard}, {Palacios}, {Charbonnel}, {Gallet}, \&
  {Bouvier}}]{Amard2016}
{Amard}, L., {Palacios}, A., {Charbonnel}, C., {Gallet}, F., \& {Bouvier}, J.
  2016, \aap, 587, A105

\bibitem[{{Bastian} \& {de Mink}(2009)}]{Bastian2009a}
{Bastian}, N. \& {de Mink}, S.~E. 2009, \mnras, 398, L11

\bibitem[{{Bastian} {et~al.}(2018){Bastian}, {Kamann}, {Cabrera-Ziri},
  {Georgy}, {Ekstr{\"o}m}, {Charbonnel}, {de Juan Ovelar}, \&
  {Usher}}]{Bastian2018a}
{Bastian}, N., {Kamann}, S., {Cabrera-Ziri}, I., {et~al.} 2018, \mnras, 480,
  3739

\bibitem[{{Bastian} {et~al.}(2016){Bastian}, {Niederhofer},
  {Kozhurina-Platais}, {Salaris}, {Larsen}, {Cabrera-Ziri}, {Cordero},
  {Ekstr{\"o}m}, {Geisler}, {Georgy}, {Hilker}, {Kacharov}, {Li}, {Mackey},
  {Mucciarelli}, \& {Platais}}]{Bastian2016a}
{Bastian}, N., {Niederhofer}, F., {Kozhurina-Platais}, V., {et~al.} 2016,
  \mnras, 460, L20

\bibitem[{{Bouvier} {et~al.}(2014){Bouvier}, {Matt}, {Mohanty}, {Scholz},
  {Stassun}, \& {Zanni}}]{Bouvier2014a}
{Bouvier}, J., {Matt}, S.~P., {Mohanty}, S., {et~al.} 2014, in Protostars and
  Planets VI, ed. H.~{Beuther}, R.~S. {Klessen}, C.~P. {Dullemond}, \&
  T.~{Henning}, 433

\bibitem[{{Brandt} \& {Huang}(2015)}]{Brandt2015a}
{Brandt}, T.~D. \& {Huang}, C.~X. 2015, \apj, 807, 25

\bibitem[{{Brott} {et~al.}(2011){Brott}, {de Mink}, {Cantiello}, {Langer}, {de
  Koter}, {Evans}, {Hunter}, {Trundle}, \& {Vink}}]{Brott2011a}
{Brott}, I., {de Mink}, S.~E., {Cantiello}, M., {et~al.} 2011, \aap, 530, A115

\bibitem[{{Charbonnel} {et~al.}(2017){Charbonnel}, {Decressin}, {Lagarde},
  {Gallet}, {Palacios}, {Auri{\`e}re}, {Konstantinova-Antova}, {Mathis},
  {Anderson}, \& {Dintrans}}]{Charbonneletal17}
{Charbonnel}, C., {Decressin}, T., {Lagarde}, N., {et~al.} 2017, \aap, 605,
  A102

\bibitem[{{Cordoni} {et~al.}(2018){Cordoni}, {Milone}, {Marino}, {Di
  Criscienzo}, {D'Antona}, {Dotter}, {Lagioia}, \& {Tailo}}]{Cordoni2018a}
{Cordoni}, G., {Milone}, A.~P., {Marino}, A.~F., {et~al.} 2018, ArXiv e-prints,
  arXiv:1811.01192

\bibitem[{{Correnti} {et~al.}(2014){Correnti}, {Goudfrooij}, {Kalirai},
  {Girardi}, {Puzia}, \& {Kerber}}]{Correnti2014a}
{Correnti}, M., {Goudfrooij}, P., {Kalirai}, J.~S., {et~al.} 2014, \apj, 793,
  121

\bibitem[{{Decressin} {et~al.}(2009){Decressin}, {Mathis}, {Palacios}, {Siess},
  {Talon}, {Charbonnel}, \& {Zahn}}]{DecressinMathis2009}
{Decressin}, T., {Mathis}, S., {Palacios}, A., {et~al.} 2009, \aap, 495, 271

\bibitem[{{Dupree} {et~al.}(2017){Dupree}, {Dotter}, {Johnson}, {Marino},
  {Milone}, {Bailey}, {Crane}, {Mateo}, \& {Olszewski}}]{Dupree2017a}
{Dupree}, A.~K., {Dotter}, A., {Johnson}, C.~I., {et~al.} 2017, \apjl, 846, L1

\bibitem[{{Eggenberger} {et~al.}(2008){Eggenberger}, {Meynet}, {Maeder},
  {Hirschi}, {Charbonnel}, {Talon}, \& {Ekstr{\"o}m}}]{Eggenberger2008a}
{Eggenberger}, P., {Meynet}, G., {Maeder}, A., {et~al.} 2008, \apss, 316, 43

\bibitem[{{Eggenberger} {et~al.}(2010){Eggenberger}, {Meynet}, {Maeder},
  {Miglio}, {Montalban}, {Carrier}, {Mathis}, {Charbonnel}, \&
  {Talon}}]{Eggenberger2010a}
{Eggenberger}, P., {Meynet}, G., {Maeder}, A., {et~al.} 2010, \aap, 519, A116

\bibitem[{{Ekstr{\"o}m} {et~al.}(2012){Ekstr{\"o}m}, {Georgy}, {Eggenberger},
  {Meynet}, {Mowlavi}, {Wyttenbach}, {Granada}, {Decressin}, {Hirschi},
  {Frischknecht}, {Charbonnel}, \& {Maeder}}]{Ekstrom2012a}
{Ekstr{\"o}m}, S., {Georgy}, C., {Eggenberger}, P., {et~al.} 2012, \aap, 537,
  A146

\bibitem[{{Espinosa Lara} \& {Rieutord}(2011)}]{EspinosaLara2011a}
{Espinosa Lara}, F. \& {Rieutord}, M. 2011, \aap, 533, A43

\bibitem[{{Georgy} {et~al.}(2013{\natexlab{a}}){Georgy}, {Ekstr{\"o}m},
  {Eggenberger}, {Meynet}, {Haemmerl{\'e}}, {Maeder}, {Granada}, {Groh},
  {Hirschi}, {Mowlavi}, {Yusof}, {Charbonnel}, {Decressin}, \&
  {Barblan}}]{Georgy2013b}
{Georgy}, C., {Ekstr{\"o}m}, S., {Eggenberger}, P., {et~al.}
  2013{\natexlab{a}}, \aap, 558, A103

\bibitem[{{Georgy} {et~al.}(2013{\natexlab{b}}){Georgy}, {Ekstr{\"o}m},
  {Granada}, {Meynet}, {Mowlavi}, {Eggenberger}, \& {Maeder}}]{Georgy2013a}
{Georgy}, C., {Ekstr{\"o}m}, S., {Granada}, A., {et~al.} 2013{\natexlab{b}},
  \aap, 553, A24

\bibitem[{{Georgy} {et~al.}(2014){Georgy}, {Granada}, {Ekstr{\"o}m}, {Meynet},
  {Anderson}, {Wyttenbach}, {Eggenberger}, \& {Maeder}}]{Georgy2014b}
{Georgy}, C., {Granada}, A., {Ekstr{\"o}m}, S., {et~al.} 2014, \aap, 566, A21

\bibitem[{{Girardi} {et~al.}(2011){Girardi}, {Eggenberger}, \&
  {Miglio}}]{Girardi2011a}
{Girardi}, L., {Eggenberger}, P., \& {Miglio}, A. 2011, \mnras, 412, L103

\bibitem[{{Goudfrooij} {et~al.}(2018){Goudfrooij}, {Girardi}, {Bellini},
  {Bressan}, {Correnti}, \& {Costa}}]{Goudfrooij2018a}
{Goudfrooij}, P., {Girardi}, L., {Bellini}, A., {et~al.} 2018, \apj, 864, L3

\bibitem[{{Goudfrooij} {et~al.}(2017){Goudfrooij}, {Girardi}, \&
  {Correnti}}]{Goudfrooij2017a}
{Goudfrooij}, P., {Girardi}, L., \& {Correnti}, M. 2017, \apj, 846, 22

\bibitem[{{Goudfrooij} {et~al.}(2014){Goudfrooij}, {Girardi},
  {Kozhurina-Platais}, {Kalirai}, {Platais}, {Puzia}, {Correnti}, {Bressan},
  {Chandar}, {Kerber}, {Marigo}, \& {Rubele}}]{Goudfrooij2014a}
{Goudfrooij}, P., {Girardi}, L., {Kozhurina-Platais}, V., {et~al.} 2014, \apj,
  797, 35

\bibitem[{{Goudfrooij} {et~al.}(2015){Goudfrooij}, {Girardi}, {Rosenfield},
  {Bressan}, {Marigo}, {Correnti}, \& {Puzia}}]{Goudfrooij2015a}
{Goudfrooij}, P., {Girardi}, L., {Rosenfield}, P., {et~al.} 2015, \mnras, 450,
  1693

\bibitem[{{Heger} \& {Langer}(2000)}]{Heger_Langer_00}
{Heger}, A. \& {Langer}, N. 2000, \apj, 544, 1016

\bibitem[{{Kamann} {et~al.}(2018){Kamann}, {Bastian}, {Husser}, {Martocchia},
  {Usher}, {den Brok}, {Dreizler}, {Kelz}, {Krajnovi{\'c}}, {Richard},
  {Steinmetz}, \& {Weilbacher}}]{Kamann2018a}
{Kamann}, S., {Bastian}, N., {Husser}, T.-O., {et~al.} 2018, \mnras, 480, 1689

\bibitem[{{Kawaler}(1988)}]{Kawaler88}
{Kawaler}, S.~D. 1988, \apj, 333, 236

\bibitem[{{Krishnamurthi} {et~al.}(1997){Krishnamurthi}, {Pinsonneault},
  {Barnes}, \& {Sofia}}]{Krishnamurthi1997a}
{Krishnamurthi}, A., {Pinsonneault}, M.~H., {Barnes}, S., \& {Sofia}, S. 1997,
  \apj, 480, 303

\bibitem[{{Lagarde} {et~al.}(2012){Lagarde}, {Decressin}, {Charbonnel},
  {Eggenberger}, {Ekstr{\"o}m}, \& {Palacios}}]{Lagarde_etal12}
{Lagarde}, N., {Decressin}, T., {Charbonnel}, C., {et~al.} 2012, \aap, 543,
  A108

\bibitem[{{Mackey} \& {Broby Nielsen}(2007)}]{Mackey2007a}
{Mackey}, A.~D. \& {Broby Nielsen}, P. 2007, \mnras, 379, 151

\bibitem[{{Maeder}(1997)}]{Maeder1997a}
{Maeder}, A. 1997, \aap, 321, 134

\bibitem[{{Maeder}(2009)}]{Maeder2009a}
{Maeder}, A. 2009, {Physics, Formation and Evolution of Rotating Stars}
  (Springer)

\bibitem[{{Maeder} \& {Meynet}(2000)}]{Maeder2000b}
{Maeder}, A. \& {Meynet}, G. 2000, Annual Review of Astronomy and Astrophysics,
  38, 143

\bibitem[{{Maeder} \& {Meynet}(2012)}]{Maeder2012a}
{Maeder}, A. \& {Meynet}, G. 2012, Reviews of Modern Physics, 84, 25

\bibitem[{{Maeder} \& {Zahn}(1998)}]{Maeder1998a}
{Maeder}, A. \& {Zahn}, J.-P. 1998, \aap, 334, 1000

\bibitem[{{Marino} {et~al.}(2018){Marino}, {Przybilla}, {Milone}, {Da Costa},
  {D'Antona}, {Dotter}, \& {Dupree}}]{Marino2018a}
{Marino}, A.~F., {Przybilla}, N., {Milone}, A.~P., {et~al.} 2018, \aj, 156, 116

\bibitem[{{Martocchia} {et~al.}(2018){Martocchia}, {Niederhofer},
  {Dalessandro}, {Bastian}, {Kacharov}, {Usher}, {Cabrera- Ziri}, {Lardo},
  {Cassisi}, {Geisler}, {Hilker}, {Hollyhead}, {Kozhurina-Platais}, {Larsen},
  {Mackey}, {Mucciarelli}, {Platais}, \& {Salaris}}]{Martocchia2018a}
{Martocchia}, S., {Niederhofer}, F., {Dalessandro}, E., {et~al.} 2018, \mnras,
  896

\bibitem[{{Mathis} {et~al.}(2018){Mathis}, {Prat}, {Amard}, {Charbonnel},
  {Palacios}, {Lagarde}, \& {Eggenberger}}]{2018arXiv180801814M}
{Mathis}, S., {Prat}, V., {Amard}, L., {et~al.} 2018, ArXiv e-prints
  [\eprint[arXiv]{1808.01814}]

\bibitem[{{Matt} {et~al.}(2015){Matt}, {Brun}, {Baraffe}, {Bouvier}, \&
  {Chabrier}}]{Mattetal2015}
{Matt}, S.~P., {Brun}, A.~S., {Baraffe}, I., {Bouvier}, J., \& {Chabrier}, G.
  2015, \apjl, 799, L23

\bibitem[{{Meynet} {et~al.}(2013){Meynet}, {Ekstr\"om}, {Maeder},
  {Eggenberger}, {Saio}, {Chomienne}, \& {Haemmerl{\'e}}}]{Meynet2013a}
{Meynet}, G., {Ekstr\"om}, S., {Maeder}, A., {et~al.} 2013, in Lecture Notes in
  Physics, Berlin Springer Verlag, Vol. 865, Studying Stellar Rotation and
  Convection, ed. M.~{Goupil}, K.~{Belkacem}, C.~{Neiner}, F.~{Ligni{\`e}res},
  \& J.~J. {Green}, 3--642

\bibitem[{{Milone} {et~al.}(2015){Milone}, {Bedin}, {Piotto}, {Marino},
  {Cassisi}, {Bellini}, {Jerjen}, {Pietrinferni}, {Aparicio}, \&
  {Rich}}]{Milone2015a}
{Milone}, A.~P., {Bedin}, L.~R., {Piotto}, G., {et~al.} 2015, \mnras, 450, 3750

\bibitem[{{Milone} {et~al.}(2017){Milone}, {Marino}, {D'Antona}, {Bedin},
  {Piotto}, {Jerjen}, {Anderson}, {Dotter}, {di Criscienzo}, \&
  {Lagioia}}]{Milone2017a}
{Milone}, A.~P., {Marino}, A.~F., {D'Antona}, F., {et~al.} 2017, \mnras, 465,
  4363

\bibitem[{{Milone} {et~al.}(2018){Milone}, {Marino}, {Di Criscienzo},
  {D'Antona}, {Bedin}, {Da Costa}, {Piotto}, {Tailo}, {Dotter}, {Angeloni},
  {Anderson}, {Jerjen}, {Li}, {Dupree}, {Granata}, {Lagioia}, {Mackey},
  {Nardiello}, \& {Vesperini}}]{Milone2018a}
{Milone}, A.~P., {Marino}, A.~F., {Di Criscienzo}, M., {et~al.} 2018, \mnras,
  477, 2640

\bibitem[{{Niederhofer} {et~al.}(2015{\natexlab{a}}){Niederhofer}, {Georgy},
  {Bastian}, \& {Ekstr{\"o}m}}]{Niederhofer2015a}
{Niederhofer}, F., {Georgy}, C., {Bastian}, N., \& {Ekstr{\"o}m}, S.
  2015{\natexlab{a}}, \mnras, 453, 2070

\bibitem[{{Niederhofer} {et~al.}(2015{\natexlab{b}}){Niederhofer}, {Hilker},
  {Bastian}, \& {Silva-Villa}}]{Niederhofer2015b}
{Niederhofer}, F., {Hilker}, M., {Bastian}, N., \& {Silva-Villa}, E.
  2015{\natexlab{b}}, \aap, 575, A62

\bibitem[{{Palacios}(2013)}]{Palacios2013}
{Palacios}, A. 2013, in EAS Publications Series, Vol.~62, EAS Publications
  Series, ed. P.~{Hennebelle} \& C.~{Charbonnel}, 227--287

\bibitem[{{Schaller} {et~al.}(1992){Schaller}, {Schaerer}, {Meynet}, \&
  {Maeder}}]{Schaller1992a}
{Schaller}, G., {Schaerer}, D., {Meynet}, G., \& {Maeder}, A. 1992, \aaps, 96,
  269

\bibitem[{{Schatzman}(1962)}]{Sch62}
{Schatzman}, E. 1962, Annales d'Astrophysique, 25, 18

\bibitem[{{Talon} \& {Charbonnel}(2004)}]{TalonCharbonnel2004}
{Talon}, S. \& {Charbonnel}, C. 2004, \aap, 418, 1051

\bibitem[{{von Zeipel}(1924)}]{vonZeipel1924a}
{von Zeipel}, H. 1924, \mnras, 84, 665

\bibitem[{{Weber} \& {Davis}(1967)}]{WD67}
{Weber}, E.~J. \& {Davis}, Jr., L. 1967, \apj, 148, 217

\bibitem[{{Yang} {et~al.}(2013){Yang}, {Bi}, {Meng}, \& {Liu}}]{Yang2013a}
{Yang}, W., {Bi}, S., {Meng}, X., \& {Liu}, Z. 2013, \apj, 776, 112

\bibitem[{{Yang} \& {Tian}(2017)}]{Yang2017a}
{Yang}, W. \& {Tian}, Z. 2017, \apj, 836, 102

\bibitem[{{Zahn}(1992)}]{Zahn1992}
{Zahn}, J.-P. 1992, \aap, 265, 115

\end{thebibliography}
\end{document}